\documentclass[runningheads]{llncs}
\usepackage[T1]{fontenc}
\usepackage{graphicx}
\usepackage{booktabs}
\usepackage{multicol}

\begin{document}
\title{Contextualizing Spotify's Audiobook List Recommendations with Descriptive Shelves}

\author{Gustavo Penha \and
Alice Wang \and 
Martin Achenbach \and 
Kristen Sheets \and
Sahitya Mantravadi \and
Remi Galvez \and
Nico Guetta-Jeanrenaud \and
Divya Narayanan \and
Ofeliya Kalaydzhyan \and
Hugues Bouchard
}
\institute{Spotify \\
\email{\{gustavop,alicew,machenbach,ksheets,sahityam,\\remigalvez,nicog,dnarayanan,ofeliyak,hb\}@spotify.com}}
\titlerunning{Contextualizing Recommendations with Descriptive Shelves}
%
%
\authorrunning{Penha et al.}
%
%
\maketitle              
\begin{abstract}

In this paper, we propose a pipeline to generate contextualized list recommendations with \emph{descriptive shelves}, in the domain of audiobooks. By creating several shelves for topics the user has an affinity to, e.g. ``\textit{Uplifting Women's Fiction}'', we can help them explore their recommendations according to their interests and at the same time recommend a diverse set of items. To do so, we use Large Language Models (LLMs) to enrich each item's metadata based on a taxonomy created for this domain. Then we create diverse descriptive shelves for each user. A/B tests show improvements in user engagement and audiobook discovery metrics, demonstrating benefits for users and content creators.
\end{abstract}

\section{Introduction}

Recommender systems help users navigate, discover, and filter through large catalogs of items. A common layout for online platforms is to have rows of recommended items grouped under a title that can be explored by scrolling horizontally. The titles of such shelves provide additional context to the items, acting as explanations to assist users in their decision-making process. 

Shelves and items contained can be personalized to each user. 
For example, one user may be predicted to have an affinity for ``\textit{Mindfulness Audiobooks}'' while another may like ``\textit{Thrilling Murder Mysteries}''. For each user, the system would recommend a shelf of contents a user will likely engage with and package the shelf with an appropriate title that describes the contents and provides context.

Previous approaches that generate content-based explanations for recommendations often rely on extracting information from user reviews, user-generated tags, or using items' rich metadata. However, such user-generated content and rich metadata are not always available. This is the case for Spotify's addition of audiobooks to the platform. To overcome this cold-start problem, we employ an extraction approach using LLMs to enrich the content of the items in the catalog with descriptors that are grounded on the item's metadata. Based on the enriched item's metadata, we propose a pipeline that generates \emph{descriptive shelves}, which group recommendations into thematic and personalized lists. The descriptive shelves aid users in their decision-making and catalog exploration process by contextualizing recommendations, while benefiting content creators by improving the discoverability and exposure of their work through a more diverse set of recommendations.



\section{Related Work}
Content-based explainability methods rely on item properties such as tags~\cite{vig2009tagsplanations}, topics~\cite{liu2024topic}, and reviews~\cite{ni2019justifying,li2021extra} to create generative~\cite{chen2021generate} or template-based~\cite{blanco2012you} explanations that go along with recommendations. By giving more context to a recommendation made by the system, explanations can achieve different goals such as improving trust, persuading users, and increasing transparency~\cite{tintarev2015explaining}. However, explaining a single recommendation is often not enough as multiple items are frequently recommended at a time on most online media platforms. When recommending sets of items we might require different explanation techniques~\cite{tintarev2012beyond,wibowo2018understanding,penha2022pairwise}. For example, Iferroudjene et. al.~\cite{iferroudjene2023methods} attempt to explain top-N recommendations for each user by identifying which previous interactions are behind the model's predictions using subgroup discovery. 

Recent advances in large language modeling~\cite{devlin2018bert,raffel2020exploring,brown2020language,dubey2024llama} have led to generative approaches addressing many recommendation problems~\cite{zhao2023recommender,deldjoo2024review,vats2024exploring} such as retrieval of candidates~\cite{rajput2023recommender}, ranking~\cite{hou2024large}, conversational recommendation~\cite{penha2020does}, and also---crucially---the generation of explanations~\cite{li2020generate,hada2021rexplug,geng2022recommendation} and enhancement of item's representations~\cite{mysore2023large}. In this paper, we test a scalable approach to explain recommendations that does not require making LLM requests for every user. We first use LLMs to extract and enhance the content-based representation we have of each entity in the catalog and use such information to explanations that contextualize lists of recommendations.

\section{Descriptive Shelves}

\subsection{Descriptor Generation}


To generate descriptors for audiobooks we first mapped out a taxonomy containing different types of descriptors. To define such taxonomy, we looked into how users search for audiobooks using both internal search queries and also requests issued at Reddit on the \textit{/r/booksuggestions/}\footnote{\url{https://www.reddit.com/r/booksuggestions/}} forum. For example, users might start a thread asking other users for ``\textit{books with a strong female protagonist}'', revealing that the characters of a book might serve as good descriptors to be used for contextualizing recommendations. This manual approach has led to the following taxonomy: (1) Genres, e.g. ``\textit{Juvenile Fiction}'', (2) Themes or topics, e.g. ``\textit{Global Politics}'', (3) Characters Descriptions, e.g. ``\textit{Female Protagonist}'', (4) Moods, e.g. ``\textit{Adventurous}'', (5) Settings, e.g. ``\textit{China's Cultural Revolution}'', (6) Personal situations, e.g. ``\textit{Dealing with Loss}'', (7) Story tropes, e.g. ``\textit{Enemies to Lovers}'', (8) Target audiences, e.g. ``\textit{Children's Literature}'', (9) Objective-based, e.g. ``\textit{Learn Japanese}'', and (10) Named entities, e.g. ``\textit{Britney Spears}''.

The metadata used as input to the LLM is the title, author(s), description, and BISAC genres\footnote{\url{https://www.bisg.org/BISAC-Subject-Codes-main}}. For each audiobook in the catalog, we use the LLM to extract the 10 types of descriptors (returning empty lists when unavailable), with a prompt containing instructions for each type of descriptor and in-context learning examples. Manual and automatic evaluations showed high accuracy in the task---by grounding the process on all the available metadata and instructing the model with the taxonomy, we minimize the risks of false positives.

\subsection{Shelf Generation}

Assuming that we have a set of descriptors for each audiobook in the catalog, we can use a recommendation model~\footnote{In our experiments we use the two-tower model described in~\cite{de2024personalized}.} to generate candidate sets $L=\{l_i, ..., l_U\}$ for each of the $U$ users that serve as input to the shelf generation approach.

\begin{figure}[h]
    \centering
    \includegraphics[width=\textwidth]{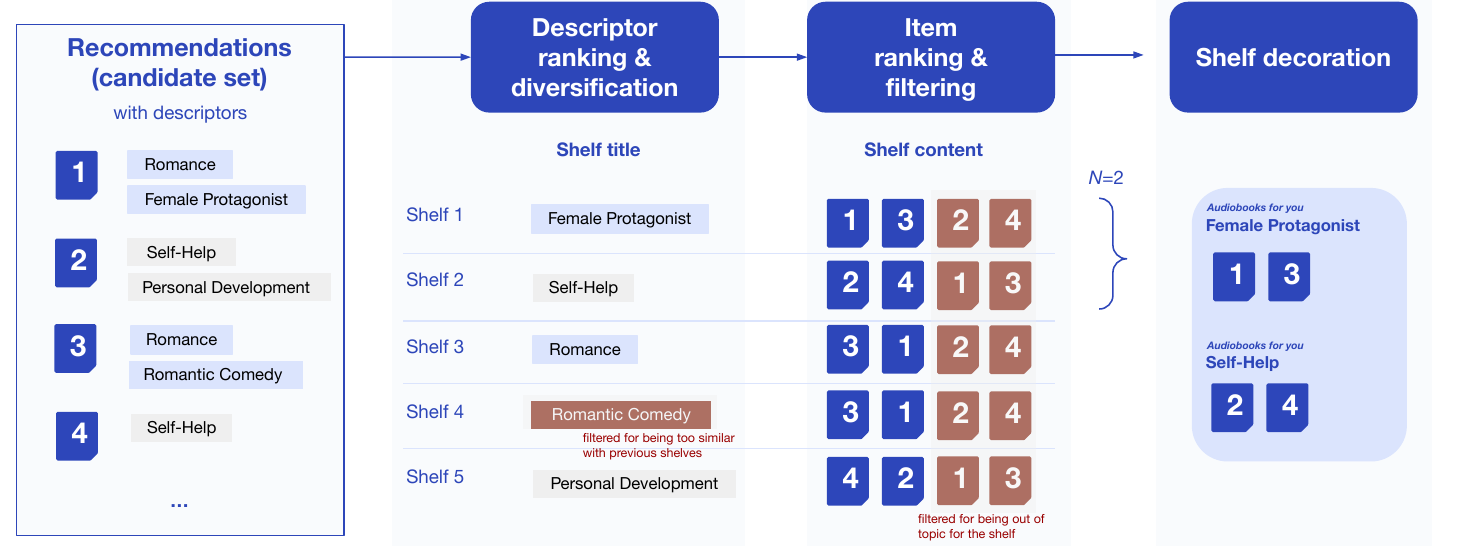}
    \caption{Shelf generation pipeline: from a set of top-K \emph{recommendations} and their respective descriptors, we \emph{rank the distinct set of descriptors} of the recommendations to be used as shelf titles for the user based on their affinities and \emph{diversify} the list by removing descriptors that are too similar to each other. Then we \emph{rank and filter the set of candidate items} for each of the distinct descriptors, and finally, we \emph{decorate} the shelves and display the top-N to the user.}
    \label{fig:pipeline}
\end{figure}

\subsubsection{Descriptor Ranking and Diversification}
The objective of these operations is to select $N$ descriptors that are diverse and relevant to the user, to serve as the titles of the shelves. For a given recommendation list $l=\{item_{1}, ..., item_{k}\}$ for user $u$, and their respective descriptors (and potential combinations\footnote{The combinations of descriptor types outlined in the taxonomy are achieved through handcrafted templates, such as \textit{<Mood>+<Genre>}, yielding combinations like “\textit{Emotional Romance}”.}) $\{D_{1}, ..., D_{k}\}$ of the types of descriptors selected, we first use a ranking function to sort the set of distinct descriptors that characterize the items in $l$ according to their relevance to the user. After that, we apply a greedy diversification approach that will filter out descriptors based on the similarity of content embeddings.

One important aspect of this stage is to define which types of descriptors from the taxonomy should be employed. Some types of descriptors such as \textit{Genres} have a smaller set of distinct descriptors that encompass larger amounts of audiobooks due to their broadness, whereas others like \textit{Themes or topics} have more fine-grained aspects of audiobooks. If we use more fine-grained and specific descriptors for shelves, we end up with narrow descriptors that might be overly specific for a user, whereas coarse-grained descriptors might not reveal enough about the audiobooks in a shelf and encompass a large part of the catalog. 

Besides the specificity of the descriptors, there is also a question on the style of the shelf title. For example, the \textit{Moods} descriptors such as \textit{Brilliant} or \textit{Powerful}, alone might not be useful to describe the audiobooks. This motivates the use of combinations of descriptors as candidates as well, such as \textit{<Mood>+<Genre>}.

\subsubsection{Item Ranking and Filtering}
The objective of these operations is to rank and filter items for each of the $N$ descriptive shelves that were selected in the previous stage. They will be used to populate the shelves and must be items that relate to the descriptors chosen as titles. The approach we use is to remove items that do not have descriptors matching the chosen descriptors and rank them according to the recommender system scores for the user.


\section{Results \& Analysis}

\paragraph{First A/B test} Firstly, we performed an A/B test on Spotify's main home surface. We replaced the existing shelf ``\textit{Audiobooks for you}'' with the descriptive shelves, e.g. ``\textit{Overcoming Obstacles Audiobooks}''. Both shelves used the same candidate set of recommendation items. The descriptive shelves had around 2500 different shelf titles, that grouped the recommendations into topics. The results revealed that the descriptive shelves increased discovery metrics by exposing users to a wider range of distinct audiobooks. However, the descriptive shelves underperformed the control cell in terms of engagement metrics. 

We hypothesize that the descriptive shelves underperformed in this test due to the following reasons. First, the interface did not indicate that the descriptive shelves are personalized (e.g. ``\textit{for you}'' suffix) due to character limitations. Second, we could only display a single shelf on the surface, and thus we have recommendations about a single topic. On the other hand, the original shelf can display recommendations of diverse topics for users. Since the majority of users in the test were cold-start users of audiobooks, a general shelf of audiobooks rather than a descriptor-specific shelf could be more effective in nudging new users toward this content type\footnote{We observed that the descriptive shelves were more successful for a subset of users with higher audiobook consumption history and narrower topics' taste.}.
\begin{table}[]
\caption{Results for the second A/B test compared to the control cell. \textit{i2c} indicates the impression to click rate and \textit{i2s} indicates the impression to stream rate. \textit{\# impressed} and \textit{\# interacted} indicates the number of distinct audiobooks impressed and interacted.}
\label{table:results}
\centering
\begin{tabular}{@{}p{3.5cm}l@{}}
\toprule
 
 & \multicolumn{1}{l}{Descriptive shelves} \\ \midrule

\textit{i2c} & 
+ 35.25\% \\
\textit{i2s} 
& + 86.96\% \\
\textit{\# impressed }
& + 627.27\% \\
\textit{\# interacted} 
& + 804.56\% \\
\bottomrule
\end{tabular}
\end{table}

\paragraph{Second A/B test}
Based on the findings of the initial test, we made some adjustments for the second A/B test. First, we changed the interface to indicate that the descriptive shelves are personalized, by adding in smaller font size ``\textit{Audiobooks for you}'' on top of the descriptive shelf name. Second, we moved the experience to Spotify's home subfeed for audiobooks, where users indicated that they had an audiobook intent by selecting the audiobook filter at the top of the home page. This allowed us to test multiple descriptive shelves, allowing the user to explore diverse topics. It also narrows down the test on users with higher intent and familiarity with audiobooks.

For this second test, we compared the shelf against audiobook shelves that were manually curated by editors, among a set of 17 different categories such as ``\textit{self-help}'', ``\textit{sports}'', and ``\textit{true crime}''. The results displayed in Table~\ref{table:results} show that descriptive shelves outperform the control cell for engagement metrics (users interact more with such shelves) and discovery metrics, (a more diverse set of audiobooks is recommended and interacted overall).

\section{Conclusion}
In this paper, we present an approach to packaging and contextualizing recommendations into descriptive shelves. The LLM helped address the data scarcity problem by producing natural language descriptions tailored to the contextualization problem. Our A/B tests reveal that the descriptive shelves can aid the user in exploring the catalog and discovering new audiobooks. By displaying descriptions that characterize the items inside each shelf, we allow users to decide whether to engage with the shelf items further (e.g. by reading the descriptions of the audiobooks) or to go to the next shelf.

In future work, we plan to explore further different types of recommender system explanations that could enhance user experience, including personalized explanations that change for each user. 
\bibliographystyle{splncs04}
\bibliography{mybibliography}
\end{document}